\documentclass[%
 aip,
 amsmath,amssymb,
 reprint,%
]{revtex4-1}

\usepackage{graphicx}
\usepackage{dcolumn}
\usepackage{bm}

\usepackage[utf8]{inputenc}
\usepackage[T1]{fontenc}
\usepackage{mathptmx}
\usepackage{etoolbox}

\makeatletter
\def\@email#1#2{%
 \endgroup
 \patchcmd{\titleblock@produce}
  {\frontmatter@RRAPformat}
  {\frontmatter@RRAPformat{\produce@RRAP{*#1\href{mailto:#2}{#2}}}\frontmatter@RRAPformat}
  {}{}
}%
\makeatother
\begin{document}

\preprint{AIP/123-QED}

\title{Periodic solution for transport of intense and coupled coasting beams through quadrupole channels}
\author{C.~Xiao}
 \altaffiliation{Author to whom correspondence should be addressed: c.xiao@gsi.de}
\author{L.~Groening}%
\affiliation{ 
GSI Helmholtzzentrum f\"ur Schwerionenforschung GmbH, D-64291 Darmstadt, Germany
}%


\date{\today}

\begin{abstract}
Imposing defined spinning to a particle beam increases its stability against perturbations from space charge~[Y.-L.~Cheon et al., Effects of beam spinning on the fourth-order particle resonance of 3D bunched beams in high-intensity linear accelerators, Phys. Rev. Accel. \& Beams {\bf 25}, 064002 (2022)]. In order to fully explore this potential, proper matching of intense coupled beams along regular lattices is mandatory. Herein, a novel procedure assuring matched transport is described and benchmarked through simulations. The concept of matched transport along periodic lattices has been extended from uncoupled beams to those with considerable coupling between the two transverse degrees of freedom. For coupled beams, matching means extension of cell-to-cell periodicity from just transverse envelopes to the coupled beam moments and to quantities being derived from these.
\end{abstract}

\maketitle

\section{Introduction}

Preservation of beam quality is of major concern for acceleration and transport especially of intense hadron beams. This aim is reached at best through provision of smooth and periodic beam envelopes, being so-called matched to the periodicity of the external focusing lattice. The latter is usually composed of a regular arrangement from solenoids or quadrupoles. For the time being, the quality of matching has been evaluated through the periodicity of spatial beam envelopes. This is fully sufficient as long as there is no coupling between the phase space planes (for brevity ``planes''), neither in beam properties nor in lattice properties.

For beams without coupling, various matching methods for intense beams have been proposed and realized in operation. First approaches, being still applied nowadays, base on differential rms-envelope equations formulated by F.~Sacherer~\cite{Sacherer,Wangler}. These assume KV-distributions and calculate space charge forces from homogeneously charged rms-equivalent ellipsoids. The forces are linear and preserve the rms-emittances. Albeit of assuming artificial KV-distributions, rms-equivalent matching of real beams has been conducted very successfully during the last decades. It became a state-of-the-art tool in operation of modern intense-beam accelerators, see~\cite{Groening_prstab2007,Groening_prl,Jeon_prab} for instance. Proper periodic solutions are especially relevant for systematic optimization of different lattice properties w.r.t. preservation of beam quality. Usually, the lattice parameter being optimized is its focusing strength, i.e., the imposed phase advance.

Variation of lattice parameters revealed many tools to optimize acceleration of intense beams with given emittances and intensity. Focusing can be accomplished by solenoids or by quadrupoles and systematic comparisons are discussed in~\cite{Reiser}. Another way is varying the phase advance along the periodic structure as considered in~\cite{Linac4}. Already in the 1960's, different quadrupole focusing schemes as FODO, FOFODODO, and FOFOFODODODO have been analyzed systematically~\cite{Friehmelt}. Recent studies revealed that imposing of spinning to the incoming beam opens another set of free parameters for further optimizing beam quality along periodic lattices~\cite{Cheon_prab}. Evidence has been provided that beam stability against perturbations from non-linear space charge forces increases with the amount of imposed spinning. This is in analogy to stability of spinning flying objects as bullets or footballs.

Spinning of beams is a very promising tool to further augment accelerator performance. It requires coupling between planes and thus imposes dedicated efforts for proper matching to periodic lattices. Beam matching with coupling between the horizontal and longitudinal planes has been investigated in~\cite{Khan_NIMa}.

The present work is on the development and demonstration of a method to assure rms-matched transport of intense beams with considerable transverse coupling, an issue being addressed conceptually in~\cite{Chernin}. It partially implements the early concept, i.e. tracking of moments, into a procedure to obtain full cell-to-cell 
four-dimensional (4D)-periodicity. Through simulations it is shown that the lattice periodicity is not just matched by the two transverse envelopes but also by the beam rms-moments that quantify coupling. To this end, an iterative procedure towards the periodic solution is applied. It starts from determining the solution with zero current, using a method that is applied later also to beams with current.

The TRACE-2D code~\cite{TRACE2D} is well suited to provide for a matching beam line between a given initial beam matrix and a desired exit beam matrix even for a full 4D scenario. However, it is an intrinsic property of the periodic-solution-problem, that the initial beam matrix at the entrance of the periodic channel is unknown. Accordingly, this code cannot be applied to the present scenario in a straight forward way.

It is explicitly stated here that providing for a specific design of the matching line itself is beyond the scope of the present work. This paper aims at demonstrating that a 4D-periodic cell-by-cell solution exists and demonstrates its derivation. Detailed definition of the specific matching line is a hard task to be addressed within future work.

The following section briefly introduces basic terms of beam rms-moments transportation through linear lattice elements. Afterwards, the beam line providing spinning, matching, and periodic focusing is introduced. The fourth section is on modeling the periodic channel for beams without and with current, followed by the description of the procedure to determine the matched solution for intense coupled beams. Finally, benchmarking of the procedure to results obtained from tracking an intense coupled Gaussian beam using a well-established simulation code is presented.
\section{Basic concepts of beam second moments transportation}
Particle coordinates are denoted by a 4$\times$1 column vector $\vec{r}\left(s\right)$ with elements $x\left(s\right)$, $x'\left(s\right)$, $y\left(s\right)$, and $y'\left(s\right)$ with
\begin{equation}
u'\left(s\right):=\frac{du\left(s\right)}{ds}\,,
\end{equation}
defining the derivation of the spatial coordinate~$u$ (refers to either $x$ or $y$) w.r.t. the longitudinal coordinate~$s$. It is assumed that the according transverse velocity~$\beta cu'$ is small in comparison to the main propagation velocity~$\beta$$c$ of the beam along~$s$. Linear transport of particle coordinates from an initial location to a final location is modeled through a linear 4$\times$4~matrix equation
\begin{equation}
\left[\vec{r}\left(s\right)\right]_{\text{final}}:=M\cdot \left[\vec{r}\left(s\right)\right]_{\text{initial}}\,.
\end{equation}

Coupled beams inhabit ten independent second-order rms-moments. They are summarized within the symmetric beam moments matrix
\begin{equation}
\label{beam_matrix}
C:=
\begin{bmatrix}
\langle xx \rangle &  \langle xx'\rangle &  \langle xy\rangle & \langle xy'\rangle \\
\langle x'x\rangle &  \langle x'x'\rangle & \langle x'y\rangle & \langle x'y'\rangle \\
\langle yx\rangle &  \langle yx'\rangle &  \langle yy\rangle & \langle yy'\rangle \\
\langle y'x\rangle &  \langle y'x'\rangle & \langle y'y\rangle & \langle y'y'\rangle
\end{bmatrix}\,.
\end{equation}

Four of its elements quantify beam coupling. Beams are $x$-$y$ coupled if at least one of these elements is different from zero. The projected rms-emittances $\varepsilon_{x}$ and $\varepsilon_{y}$ are defined through the determinants of the two on-diagonal sub-matrices as
\begin{equation}
\label{projected_emittances}
\varepsilon_{u}=\sqrt{\langle uu\rangle\langle u'u'\rangle-\langle uu'\rangle^2}\,,
\end{equation}
i.e, they do not depend on coupled beam moments. In turn, the two eigen-emittances
\begin{equation}
\label{eigen_emittances}
\varepsilon_{1,2}=\frac{1}{2}\sqrt{-\text{tr}\left(CJ\right)^2\pm\sqrt{\text{tr}^2\left(CJ\right)^2-16\det\left({C}\right)}}\,,
\end{equation}
depend on all beam moments including those with coupling. Any linear transformation~$M$ obeying
\begin{equation}
\label{beam_matrix_J}
J=M^\text{T}\cdot J \cdot M\,,~~~~~~
	J:=
\begin{bmatrix}
		-1 & 0 &0 & 0\\
        0 & 0& 0& 1\\
        0 &0 & -1& 0\\
\end{bmatrix}\,,
\end{equation}
is called symplectic and it preserves the two eigen-emittances. Just if~$M$ does not include any coupling elements, it will also preserve the two projected rms-emittances. In case that a transformation~$M$ decouples a given beam, the decoupled beam's rms-emittances are equal to the two eigen-emittances which remained unchanged by~$M$. Coupling can be quantified by the coupling parameter~\cite{ROSE}
\begin{equation}
	t:=\frac{\varepsilon_x\varepsilon_y}{\varepsilon_1\varepsilon_2}-1\,,~~~~~~\varepsilon_{4\text{d}}=\varepsilon_1\varepsilon_2\,,
\end{equation}
and if and only if $t$ is equal to zero, there is no inter-plane correlation and the projected rms-emittances are equal to the eigen-emittances.

A simple way to impose spinning to a beam is to pass it through an effective half solenoid. Although half solenoids do not exist due to $\vec{\nabla}\cdot\vec{B}=0$, their effect can be imposed by particle creation inside the solenoid or by changing the beam charge state inside the solenoid. The first method is applied in Electron-Cyclotron-Resonance (ECR) ion sources~\cite{Betrand,Xiao_nim2013} and the second method has been proposed in~\cite{Groening_prstab2011,Xiao_prstab2013} and demonstrated experimentally in~\cite{Groening_prl2014}. Effective half solenoids have the appealing feature that decoupling afterwards for further transportation is quasi independent from their magnetic field strength~\cite{Xiao_prstab2013,Groening_arxiv}.

The first part of the transport matrix~$S^h$ of an effective half solenoid is given by the matrix~$S_{\rightarrow}$ of the main body of the solenoid of effective length~$L$, comprising just the pure longitudinal magnetic field~$B_s$
\begin{equation}
\label{beam_matrix_sol_body}
	S_{\rightarrow}=
\begin{bmatrix}
		1 &  \frac{\sin\left(2KL\right)}{2K}& 0 & \frac{1-\cos\left(2KL\right)}{2K}\\
		0 & \cos\left(2KL\right) &0 & \sin\left(2KL\right)\\
        0 & -\frac{1-\cos\left(2L\right)}{2K}& 1& \frac{\sin\left(2KL\right)}{2K}\\
        0 &-\sin\left(2KL\right) & 0& \cos\left(2KL\right)\\
\end{bmatrix}\,,
\end{equation}
with~$K:=B_s/\left[2\left(B\rho\right)\right]$ (Larmor wave number) and $\left(B\rho\right)$ as beam rigidity. $K$ imposes spinning to the particles through preservation of the canonical angular momentum during transition through the solenoid, i.e.,
\begin{equation}
L_{\theta}=m\gamma r v_{\theta}+\frac{qB_{s}}{2}r^2=\text{const}\,
\end{equation}
being also known as Busch's theorem~\cite{Busch}. Assuming an incoming particle without canonical angular momentum in front of the solenoid with
\begin{equation}
v_{\theta}=0\,,~~~~~~L_{\theta}=0\,,
\end{equation}
inside the solenoid $v_{\theta}$ will be changed to 
\begin{equation}
v_{\theta}=-\frac{qB_s}{2\gamma m}r\,,
\end{equation}
being equivalent to introduction of spinning by the solenoid. The extension of Busch's theorem from one single particle to beams is treated in~\cite{Busch2}. 

The second part of~$S^h$ is from the fringe field matrix~$S_{\downarrow}$ of the solenoid exit
\begin{equation}
\label{beam_matrix_sol_exit}
	S_{\downarrow}=
\begin{bmatrix}
		1 &  0 & 0 & 0\\
		0 & 1 & -K & 0\\
        0 & 0& 1& 0\\
        K &0 & 0& 1\\
\end{bmatrix}\,,
\end{equation}
and the total matrix of the half solenoid is the product of both matrices
\begin{equation}
S^h=S_{\downarrow}\cdot S_{\rightarrow}=
\begin{bmatrix}
		S^h_{xx} & S^h_{xy}\\
		S^h_{yx} &  S^h_{yy}\\
\end{bmatrix}\,.
\end{equation}
The total transfer matrix $S$ of a complete solenoid is the product of entrance matrix $S_{\uparrow}$, main body matrix~$S_{\rightarrow}$, and exit matrix~$S_{\downarrow}$
\begin{equation}
S=S_{\downarrow}\cdot S_{\rightarrow}\cdot S_{\uparrow}\,,~~~~~~S_{\uparrow}=
\begin{bmatrix}
		1 &  0 & 0 & 0\\
		0 & 1 & K & 0\\
        0 & 0& 1& 0\\
        -K &0 & 0& 1\\
\end{bmatrix}\,.
\end{equation}
The determinants of the diagonal sub-matrices~$S^h_{xx}$ and~$S^h_{yy}$ are different from~1.0, hence the projected rms-emittances are changed by~$S^h$. Additionally, $S^h_{xy}$ and~$S^h_{yx}$ are also different from zero, thus coupling will be imposed to an initially uncoupled beam. Although being non-symplectic, $S^h$ has the determinant of~1.0 preserving the product of the two eigen-emittances. 

The sub-matrices of a transport matrix~$Q$ of a regular quadrupole of strength~$k:=Gl/\left(B\rho\right)$ and effective length~$l$ are given by
\begin{equation}
\label{beam_matrix_qua_1}
	Q_{xx}=
\begin{bmatrix}
		\cos\left(kl\right) &  \frac{\sin\left(kl\right)}{k}\\
		-k\sin\left(kl\right) & \cos\left(kl\right) \\
\end{bmatrix}\,,
\end{equation}
and
\begin{equation}
\label{beam_matrix_qua_2}
	Q_{yy}=
\begin{bmatrix}
		\cosh\left(kl\right) & \frac{\sinh\left(kl\right)}{k}\\
		k\sinh\left({kl}\right) & \cosh\left(kl\right) \\
\end{bmatrix}\,,
\end{equation}
with $G$ being the magnetic field gradient of the quadrupole implying \mbox{$B_y=Gx$} and $B_x=-Gy$. For positive (negative)~$G$, quadrupoles focus in the horizontal (vertical) plane and defocus in the vertical (horizontal) plane. The coupling sub-matrices are zero. The matrix of a drift is  
\begin{equation}
\label{beam_matrix_qua_3}
	D_{xx}=D_{yy}=
	\begin{bmatrix}
		1 & l\\
		0 & 1\\
	\end{bmatrix}\,,
\end{equation}
with its sub-matrices being equal to zero. Finally, clockwise rotation of the beam by~$\theta$ around the positive~$s$-axis is modeled through the symplectic matrix
\begin{equation}
\label{rotation}
	R\left(\theta\right)=
	\begin{bmatrix}
		\cos\left(\theta\right) &  0 & -\sin\left(\theta\right) & 0\\
		0 & \cos\left(\theta\right) & 0 & -\sin\left(\theta\right)\\
        \sin\left(\theta\right) & 0& \cos\left(\theta\right)& 0\\
        0 & \sin\left(\theta\right) & 0& \cos\left(\theta\right)\\
	\end{bmatrix}\,.
\end{equation}
\section{Beam line for coupling, matching, and transportation}
\begin{figure}[hbt]
	\centering
	\includegraphics*[width=85mm,clip=]{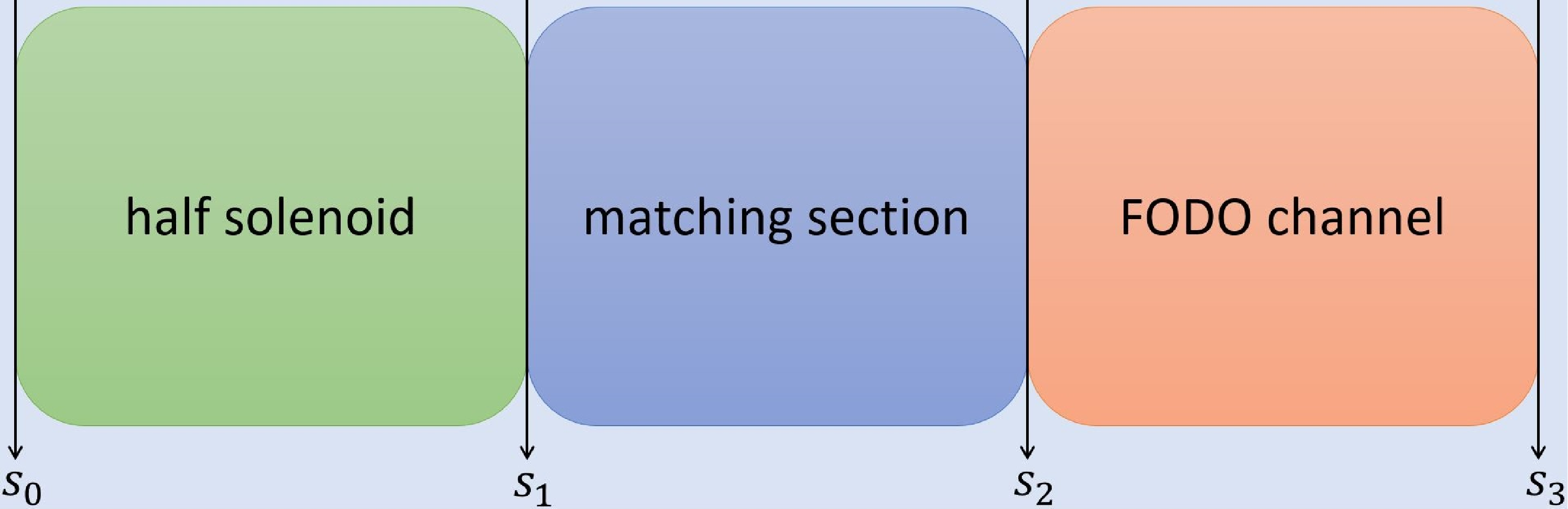}
	\caption{The beam line comprises three parts: ($\text{\MakeUppercase{\romannumeral1}}$) effective half solenoid; ($\text{\MakeUppercase{\romannumeral2}}$) matching section; ($\text{\MakeUppercase{\romannumeral3}}$) regular quadrupole doublet section (twelve cells). Space charge effects are not considered along the first two sections~(see text).}
	\label{transfer_line}
\end{figure}

The beam line being used to determine the periodic solution of an intense coupled beam along a periodic channel is sketched systematically in Fig.~\ref{transfer_line}. It comprises three sections, starting with an effective half solenoid being followed by a matching section. This section transports the beam parameters from the solenoid exit to the entrance of the periodic channel. These two sections include coupling elements. The third section is a periodic sequence of non-coupling regular quadrupoles. 

At the beginning of the beam line, an uncoupled beam is assumed with beam sigma-matrix
\begin{equation}
\label{beam_matrix_1}
	C\left(s_0\right)=
	\begin{bmatrix}
		C_{xx} &  O \\
		O & C_{yy} \\
	\end{bmatrix}\,,
\end{equation}
\begin{equation}
\label{beam_matrix_2}
	C_{xx}=
	\varepsilon_{x}
\begin{bmatrix}
		\beta_{x} &  -\alpha_{x} \\
		-\alpha_{x} & \frac{1+\alpha_{x}^2}{\beta_{x}} \\
	\end{bmatrix}\,,~
	C_{yy}=
	\varepsilon_{y}
\begin{bmatrix}
		\beta_{y} &  -\alpha_{y} \\
		-\alpha_{y} & \frac{1+\alpha_{y}^2}{\beta_{y}} \\
	\end{bmatrix}\,,
\end{equation}
with $\beta _{u}$~=~$\langle uu\rangle$/$\varepsilon _{u}$ and $\alpha _{u}$~=~$-\langle uu'\rangle$/$\varepsilon _{u}$. The beam matrix at the beginning of the matching section is
\begin{equation}
\label{half_sol}
C\left(s_1\right)=S^h \cdot C\left(s_0\right) \cdot \left({S^h}\right)^\text{T}\,.
\end{equation}

\begin{figure}[hbt]
	\centering
	\includegraphics*[width=85mm,clip=]{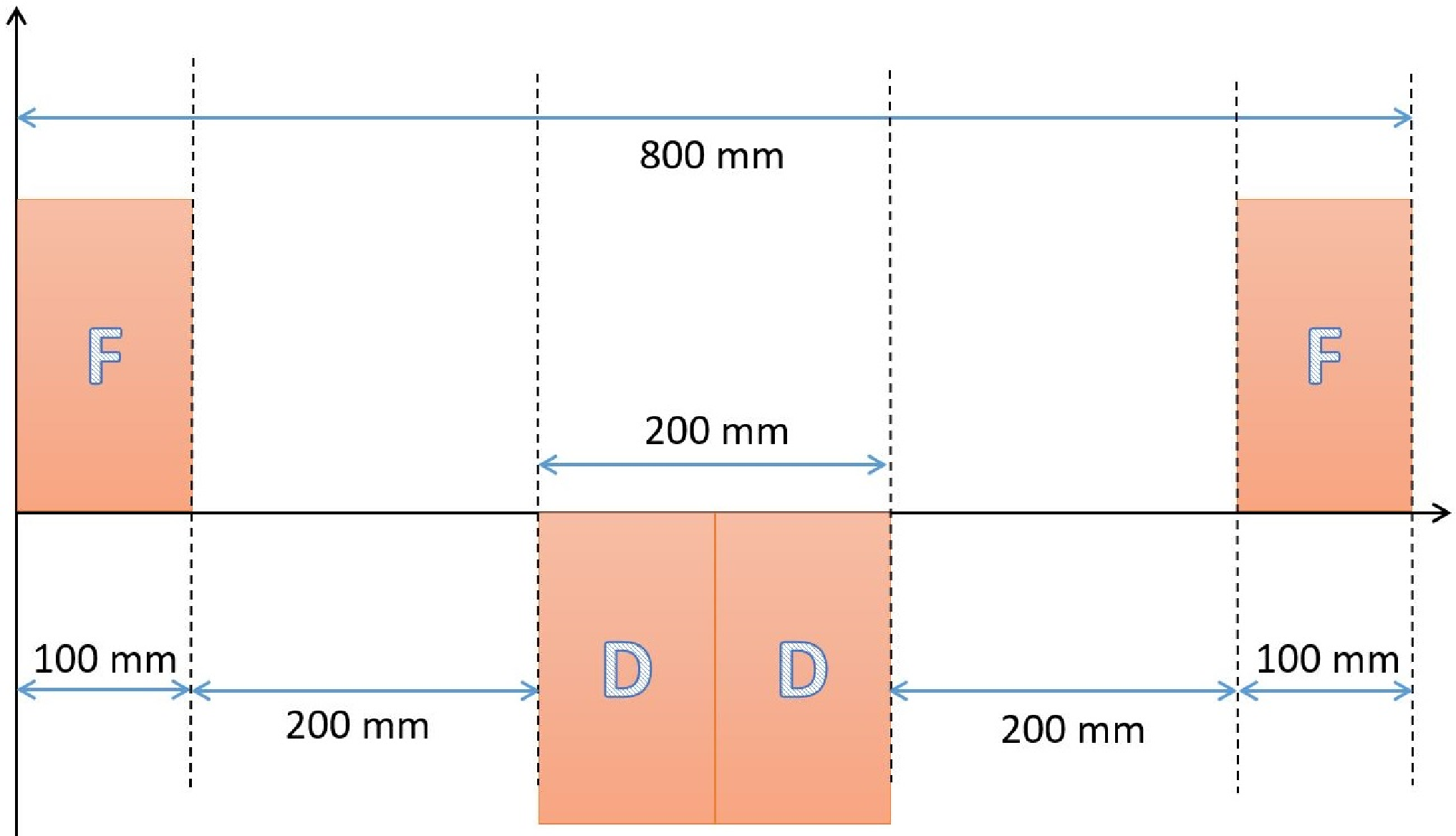}
	\caption{One cell of the periodic quadrupole channel (cell length $\ell$~=~0.8~m). Focusing and defocusing quadrupoles have gradients of $G=\pm$1.0~T/m.}
	\label{FODO}
\end{figure}

The matching section is modeled through the symplectic and coupling matrix~$\Re$ and hence
\begin{equation}
C\left(s_2\right)=\Re\cdot C\left(s_1\right)\cdot {\Re}^\text{T}\,,
\end{equation}
is the beam matrix at the entrance to the quadrupole channel. Figure~\ref{FODO} depicts one cell of the quadrupole FODO channel

Its transport matrix is a product of five single matrices
\begin{equation}
\Im=Q^{h}_{f}\cdot D\cdot Q_{d}\cdot D\cdot Q^{h}_{f}\,,~~~~~~Q_{f}=Q_{f}^{h}\cdot Q^{h}_{f}\,.
\end{equation}

\begin{figure}[hbt]
	\centering
	\includegraphics*[width=85mm,clip=]{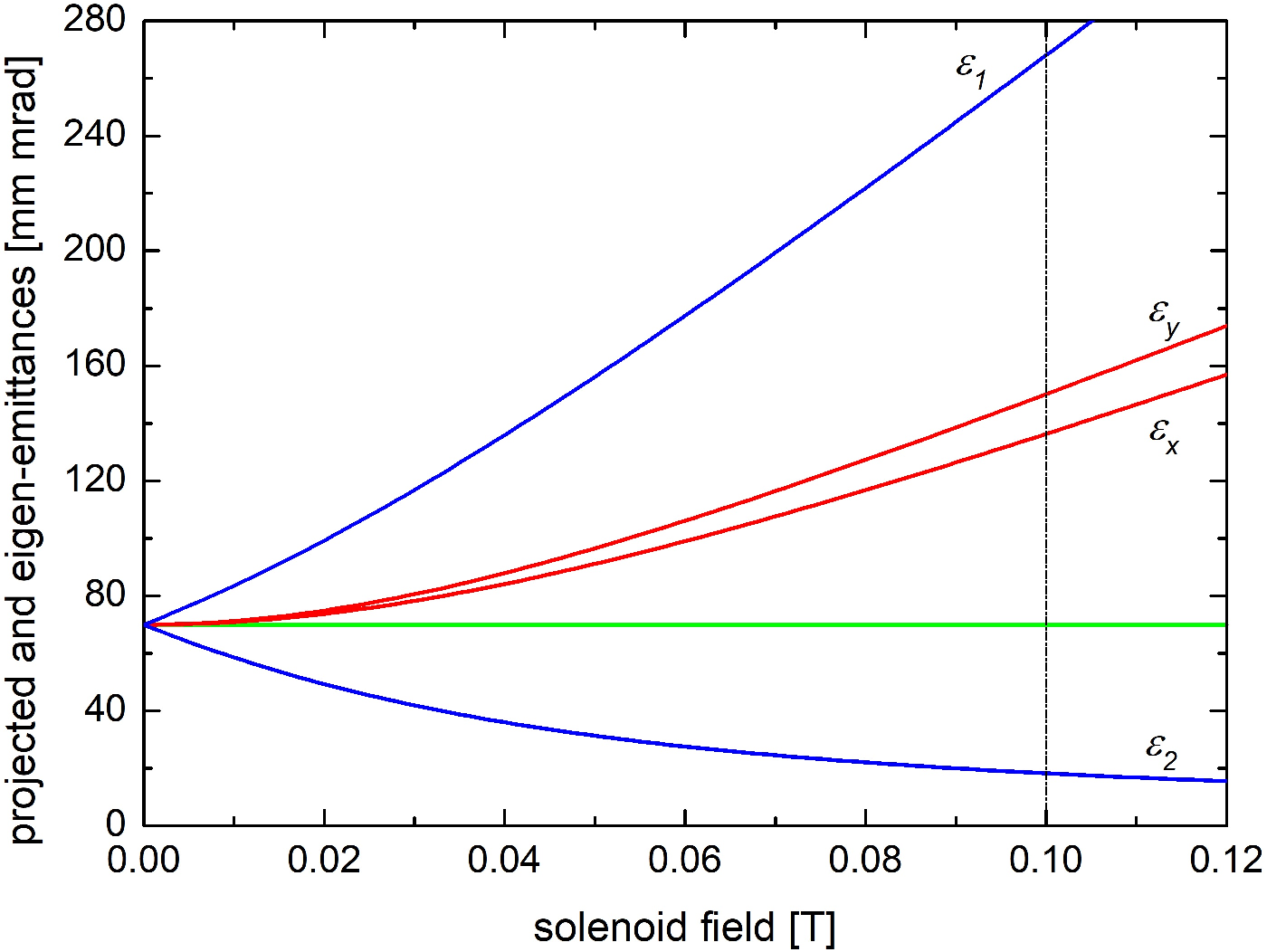}
	\caption{Projected rms-emittances (red), eigen-emittances (blue), and square root of 4d-emittance (green) at the exit of half solenoid. If the solenoid field is off, $\varepsilon_{x}$~=~$\varepsilon_{y}$~=~$\varepsilon_{1}$~=~$\varepsilon_{2}$=~69.90~mm~mrad. If the solenoid field is $B_s$~=~0.1~T, $\varepsilon_{x}$~=~136.3, $\varepsilon_{y}$~=~150.1, $\varepsilon_{1}$~=~268.0, and $\varepsilon_{2}$~=~18.23~mm~mrad with coupling factor of $t$~=~3.186.}
	\label{sol_field}
\end{figure}

At the entrance to the beam line at~$s_0$, an uncoupled proton beam with an energy of 150~keV/u is assumed. The type of channel corresponds to a common scheme of focusing. Low energy protons at this energy are provided at many sources around the world. The rigidity allows to apply solenoid field strengths being reasonably low in order to provide for a considerable amount of coupling. Beam Twiss parameters are set to $\varepsilon_{x}$~=~$\varepsilon_{y}$~=~69.90~mm~mrad, $\beta_{x}$~=~$\beta_{y}$~=~2~m/rad, $\alpha_{x}$~=~0.250, and $\alpha_{y}$~=~-0.275. If the length of the half solenoid is set to~0.25~m, the values of eigen-emittances and projected emittances at the exit of the half solenoid (position $s_1$) are determined by the solenoid field as shown in~Fig.~\ref{sol_field}.

After transport through this half solenoid the beam matrix (in units of mm and mrad) is
\begin{equation}
C\left(s_1\right)=
\begin{bmatrix}
		+133.6 & -8.578&+2.021 &+124.9  \\
         -8.578 & +139.5 &-124.9& -31.08 \\
        +2.021 & -124.9 &+151.4 & +28.22 \\
        +124.9 &  -31.08 &+28.22 & +154.1 \\
\end{bmatrix}\,,
\end{equation}
in order to obtain a periodic solution for this coupled beam, the details of the matching section are not required as seen in the following. However, it is modeled by a transport matrix including 16 elements 
\begin{equation}
\Re\left(m_1,m_2,\dots,m_{16}\right)=
\begin{bmatrix}
		m_1 & m_2 &m_3 &m_4  \\
        m_5 & m_6 &m_7 &m_8  \\
        m_9 & m_{10} &m_{11} &m_{12}  \\
        m_{13} & m_{14} &m_{15} &m_{16} \\
	\end{bmatrix}\,.
\end{equation}

Although initially being unknown, the 16 elements must provide for $\det\left({\Re}\right)$~=~1.0 and that $\Re$ is symplectic according to~Eq.~(\ref{beam_matrix_J}). For brevity, the set of $m_1,m_2,\dots,m_{16}$ shall be denoted by~$\aleph$. Although the detailed layout of the matching section is beyond the scope of this paper, a conceptual approach is sketched in the~Appendix~C.

\section{Modeling of periodic channel}
For zero current, the effective focusing forces are given solely by the external lattice. The actual beam shape has no influence on them and therefore the periodic solution even for coupled beams may be found analytically. For intense beams instead, defocusing space charge forces depend on the beam shape and orientation in real space. Actually, they depend also on the spatial distribution. However, since modeling of space charge forces using rms-equivalent KV-distributions proofed to work very well for matching purposes, this approach is followed here as well. 

In the following, an iterative method is described to determine the periodic solution for zero current. At first glance, it seems more complicated w.r.t. a straight analytical approach. However, it has the advantage to be applicable easily to obtain the periodic solution even with current.
\subsection{beam with zero current}
\label{subsec_4a}
The periodic solution meets the condition
\begin{equation}
\label{matching_FODO}
C\left(s_2\right)=\Im\cdot C\left(s_2\right)\cdot \Im^\text{T}=C\left(s_2+\ell\right)\,,
\end{equation}
where $\ell$ is the length of one cell and the transport matrix from the exit of the solenoid~$s_1$ to the exit of the first cell is 
\begin{equation}
\mho\left(\aleph\right)=\Im\cdot \Re\left(\aleph\right)\,,
\end{equation}
where $\Im$ is fully known from the cell of the quadrupole channel~(see Fig.~\ref{FODO}).

From first principles, neither the periodic solution is known nor are the elements~$\aleph$ that provide for the according matching from the exit of the solenoid~$s_1$ to the entrance of the channel~$s_2$. The iterative procedure to obtain finally both, starts with a guessed initial set~$\aleph^{i}$ that just meets the condition of being symplectic and~$\det{\left[{\Re\left({\aleph}^{i}\right)}\right]}$~=~1.0. It will most likely not meet the condition of the periodic solution, i.e.,
\begin{equation}
\label{zero_current_mismatching}
\Re\left(\aleph^{i}\right)\cdot C\left(s_1\right)\cdot\Re^{\text{T}}\left(\aleph^{i}\right)\neq \mho\left(\aleph^{i}\right)\cdot C\left(s_1\right)\cdot\mho^{\text{T}}\left(\aleph^{i}\right)\,,
\end{equation}
hence the beam matrix in front of the channel is different from the one behind the first cell (see details in appendix~A).

With the MATHCAD~\cite{mathcad} routine $\it Minerr$, a set of matching matrix elements $\aleph^{0}$ for zero beam current can be found, such that the symplectic condition and $\text{det}\left[\Re\left(\aleph^{0}\right)\right]$~=~1.0 is met sharply together with providing periodicity. The routine is dedicated to solve an under-determined system of equations with a defined set boundary conditions.
\begin{equation}
\label{zero_current_matching}
\Re\left(\aleph^{0}\right)\cdot C\left(s_1\right)\cdot\Re^{\text{T}}\left(\aleph^{0}\right)=\mho\left(\aleph^{0}\right)\cdot C\left(s_1\right)\cdot\mho^{\text{T}}\left(\aleph^{0}\right)\,.
\end{equation}

With $\aleph^{0}$ being determined, the periodic beam matrix at the beginning of the channel has been calculated as
\begin{equation}
C^{0}\left(s_2\right)=
\begin{bmatrix}
		+158.1 & +0.000 &-76.88 &+95.30 \\
        +0.000 & +97.93 &-27.65 & -164.1 \\
        +76.88 & -27.65 &+56.66 &+0.000 \\
        +95.30 &  -164.1 &+0.000& +438.9 \\
\end{bmatrix}\,,
\end{equation}
and it is equal to $C^{0}\left(s_2+\ell\right)$. As for the case of an uncoupled beam, the periodic solution of the coupled beam features~$\alpha_{x,y}$~=~0 as expected from the symmetry of the regular cell of the channel. However, the corresponding coupling parameters from combinations of other planes are different from zero due to inter-plane coupling. The zero current transport matrix of one cell (in units of m and rad) is
\begin{equation}
\Im\left(\aleph^{0}\right)=
\begin{bmatrix}
		+0.321 & +1.203&+0.000&+0.000\\
        -0.745 & +0.321&+0.000&+0.000\\
       +0.000&+0.000&+0.321&+0.349\\
      +0.000&+0.000&-2.569&+0.321\\
\end{bmatrix}\,,
\end{equation}
and evaluation of its sub-traces delivers the zero current phase advance of $\mu_0$~=~71.26$^{\circ}$. The zero current transport matrix $\Im\left(\aleph^{0}\right)$ is independent of the initial beam matrix $C^{0}\left(s_2\right)$ and is determined only by the lattice of the quadrupole channel.
\subsection{beam with high current}
\label{subsec_4b}
\begin{figure}[hbt]
	\centering
	\includegraphics*[width=45mm,clip=]{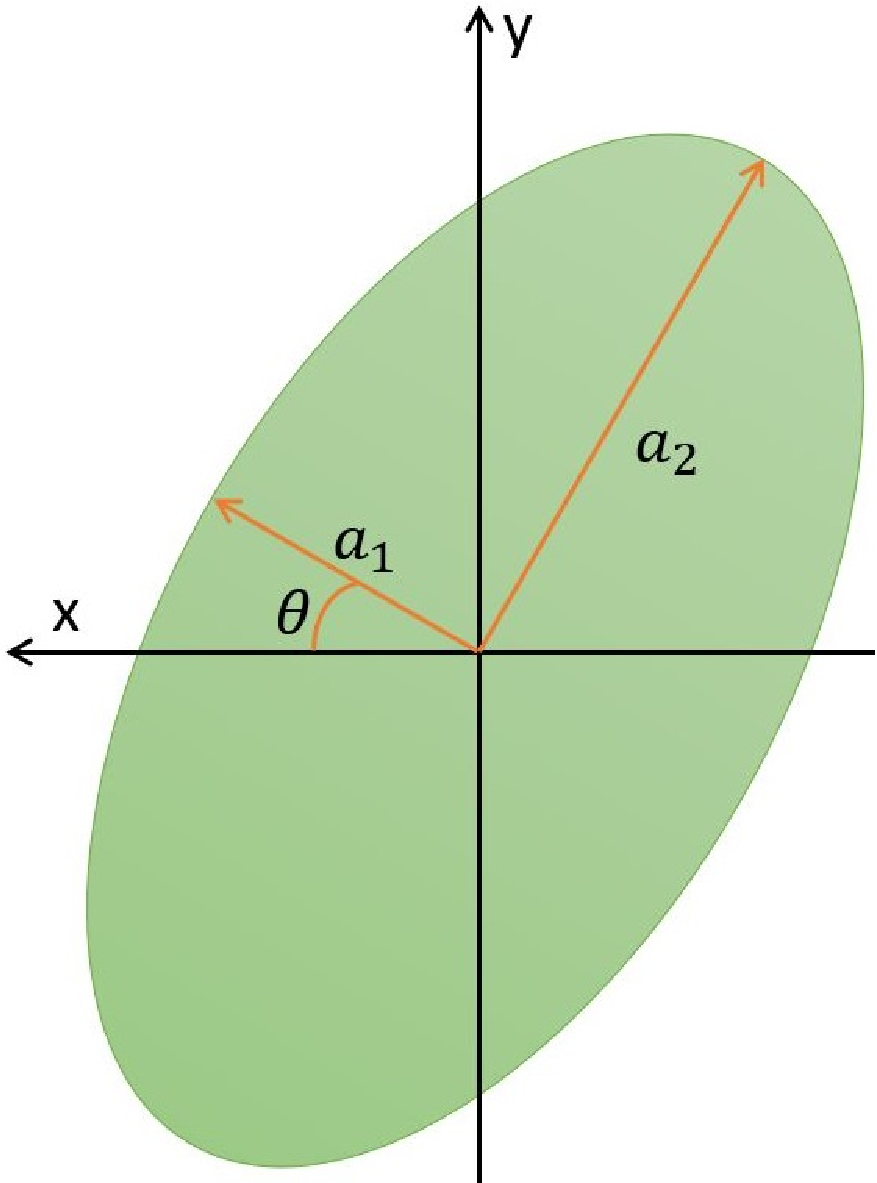}
	\caption{Ellipse of an $x$-$y$ coupled beam in real space. $A_{xy}$ is the rms-area of the beam, see Eq.~(\ref{emittance_xy_1}). Parameters $\alpha_{xy}$ and $\beta_{xy}$ are its equivalent Twiss parameters defining the ellipse orientation and aspect ratio in real space. The $x$, $y$, and $s$ unit vectors of the Cartesian coordinate system follow the right-hand rule.}
	\label{coupled_ellipse}
\end{figure}

For KV-beams, the electric self-field caused by space charge can be calculated analytically as done by Sacherer~\cite{Sacherer} for uncoupled beams, i.e., for upright ellipses. In case of coupling, the ellipse is generally tilted as drawn in Fig.~\ref{coupled_ellipse}. Here, the space charge forces are firstly calculated within the tilted frame. In a second step, these forces are projected into the upright laboratory frame and applied to the beam. They are equivalent to a defocusing quadrupole kick in both planes. The strengths are not equal along both planes but the resulting 4D-transformation is linear and symplectic. Hence it will be modeled by another 4$\times$4 transport matrix~$\varkappa$.

The ellipse is described by its two semi-axes $a_1$ and $a_2$ and by the rotation angle $\theta$ of $a_1$ w.r.t. $x$-axis. Its rms-area is given by
\begin{equation}
\label{emittance_xy_1}
A_{xy}=\sqrt{\langle xx\rangle \langle yy\rangle-\langle xy\rangle ^2}=a_1a_2\,.
\end{equation}

The above ellipse parameters are calculated from the beam second moments through
\begin{equation}
	\label{Twiss_xy_1}
	\beta_{xy}=\frac{\langle xx\rangle}{A_{xy}}\,,~~~~~~
	\alpha_{xy}=-\frac{\langle xy\rangle}{A_{xy}}\,,
\end{equation}
\begin{equation}
\Theta=\frac{1}{2}\arctan\frac{-2\alpha_{xy}}{\beta_{xy}-\frac{1+\alpha^2_{xy}}{\beta_{xy}}}\,,~~~~
h=\frac{\beta_{xy}}{2}+\frac{1+\alpha^2_{xy}}{2\beta_{xy}}\,,
\end{equation}
and
\begin{equation}
a_{1,2}=\sqrt{\frac{A_{xy}}{2}}\left(\sqrt{h+1}\pm\sqrt{h-1}\right)\,.
\end{equation}

The transport matrix $\varkappa$ is calculated from the ellipse geometric parameters and the general beam parameters as
\begin{equation}
\varkappa=R^{-1}\left(\Theta\right)\cdot \varkappa^{\ast}\cdot R\left(\Theta\right)\,,
\end{equation}
where $\varkappa^{\ast}$ is the matrix in the tilted ellipse frame. It reads 
\begin{equation}
	\varkappa_{1,2}^{\ast}=
\begin{bmatrix}
		1 & 0 \\
		\kappa_{1,2} \delta s & 1\\
\end{bmatrix}
	\,,~~~~~~
	\varkappa^{\ast}=
\begin{bmatrix}
		\varkappa_{1}^{\ast} & O \\
		O & \varkappa_{2}^{\ast}\\
\end{bmatrix}
	\,,
\end{equation}
with $\delta s$ being the step size along $s$ between two space charge kicks. $\kappa_{1,2}$ are the respective kick strengths along each semi-axis and are given by
\begin{equation}
\kappa_1=\frac{\kappa_{\text{sc}}}{2a_1\left(a_1+a_2\right)}\,,~~~~~~\kappa_2=\frac{\kappa_{\text{sc}}}{2a_2\left(a_1+a_2\right)}\,,
\end{equation}
from the generalized beam perveance
\begin{equation}
\kappa_{\text{sc}}=\frac{qI}{2\pi\epsilon_0 m\left(\gamma\beta c\right)^3}\,,
\end{equation}
with $q$ as particle charge, $I$ as beam current, and $\beta$ and $\gamma$ as relativistic factors.

With these prerequisites, any beam line from (skewed) quadrupoles transporting a coupled intense beam is modeled through a sequence of symplectic linear transport matrices. Quadrupoles and drifts are sub-divided into many slices each and transportation through them is by a sequence of transports along slice length $\delta s$ without space charge and execution of the space charge kick with $\varkappa$ afterwards. This method has been implemented into many codes. For uncoupled beams, the PARMILA code~\cite{PARMILA} for instance uses it to design periodic lattices and to evaluate their performances. Here it shall serve to obtain cell-by-cell periodic solutions for intense coupled beams.
\section{Periodic solution with space charge and coupling}
Solutions of the beam matrix along the periodic channel are considered as periodic, if the equation
\begin{equation}
C\left(s_2\right)\approx C\left({s_2+\ell}\right)\,
\end{equation}
is fulfilled to very good approximation. Subsection~\ref{subsec_4a} presented such a solution $C^0\left(s_2\right)$ for zero current. This solution will not hold with beam current being switched on. This is from the dependence of the cell transport matrix $\Im$ from the beam current and from the beam Twiss parameters at the entrance to the channel as shown in subsection~\ref{subsec_4b}.

In order to find a solution that holds even with current, another iterative procedure is applied. It uses the method of determining a matching setting $\aleph$ presented in section~\ref{subsec_4a}. Additionally, it performs an iterative switching between obtaining the periodic transport matrix from tracking and using it to re-adapt the matching to it.

The iterative procedure starts from the beam moments matrix $C\left(s_1\right)$ behind the solenoid being then transported through the matching line $\Re\left(\aleph^{0}\right)$ for zero current. The resulting beam matrix at the entrance to the channel

\begin{equation}
C^{0}\left(s_2\right)=\Re\left(\aleph^{0}\right)\cdot C\left(s_1\right) \cdot \Re^{\text{T}}\left(\aleph^{0}\right)\,,
\end{equation}
is then tracked with high current (10~mA) through one cell. Accordingly, the total transport matrix of the cell $\Im_{\text{sc}}\left(\aleph^{0}\right)$ is a result of the tracking procedure described in subsection~\ref{subsec_4b}. $\Im_{\text{sc}}\left(\aleph^{0}\right)$ depends on the current $I$ and on the spatial beam parameters at the entrance of the channel. The 4$\times$4 elements of $\Im_{\text{sc}}\left(\aleph^{0}\right)$ are stored for further use. Most likely, $C^{0}\left(s_2\right)$ does not meet the condition of the periodic solution with current, i.e,
\begin{equation}
C^{0}\left(s_2\right)\neq \Im_{\text{sc}}\left(\aleph^{0}\right)\cdot\Re\left(\aleph^{0}\right)\cdot C\left(s_1\right)\cdot\Re^{\text{T}}\left(\aleph^{0}\right)\cdot\Im^{\text{T}}_{\text{sc}}\left(\aleph^{0}\right)\,.
\end{equation}

However, the cell matrix $\Im_{\text{sc}}\left(\aleph^{0}\right)$ is used to re-adapt the matching setting such, that a new matching $\aleph^1$ is found which provides for equal beam matrices before and after transport through the cell matrix~$\Im_{\text{sc}}\left(\aleph^{0}\right)$
\begin{equation}
C^{1}\left(s_2\right)=\Im_{\text{sc}}\left(\aleph^{0}\right)\cdot\Re\left(\aleph^{1}\right)\cdot C\left(s_1\right)\cdot\Re^{\text{T}}\left(\aleph^{1}\right)\cdot\Im^{\text{T}}_{\text{sc}}\left(\aleph^{0}\right)\,,
\end{equation}
emphasizing that the above equation uses the stored elements of~$\Im_{\text{sc}}\left(\aleph^{0}\right)$.

This new matching $\aleph^{1}$ delivers the beam matrix $C^{1}\left(s_2\right)$ in front of the channel. It is now re-tracked with current through the cell as described in subsection~\ref{subsec_4b}. The tracking will provide a new cell matrix $\Im_{\text{sc}}\left(\aleph^{1}\right)$. Again its 4$\times$4 elements are stored to re-adapt the matching to a setting $\aleph^{2}$ meeting the periodic solution assuming the new matrix $\Im_{\text{sc}}\left(\aleph^{1}\right)$ along the channel
\begin{equation}
C^{2}\left(s_2\right)=\Im_{\text{sc}}\left(\aleph^{1}\right)\cdot\Re\left(\aleph^{2}\right)\cdot C\left(s_1\right) \cdot\Re^{\text{T}}\left(\aleph^{2}\right)\cdot\Im^{\text{T}}_{\text{sc}}\left(\aleph^{1}\right)\,.
\end{equation}

This in turn provides a new beam matrix $C^{2}\left(s_2\right)$ in front of the channel, which changes the transport matrix of the cell to~$\Im_{\text{sc}}\left(\aleph^{2}\right)$. Continuing this procedure finally converges, i.e., the changes from $\aleph^{n-1}$ to $\aleph^{n}$ become very small and finally negligible. Accordingly, after a sufficient amount of iterations~$\j$, the periodic condition is fulfilled through
\begin{equation}
\label{high_current_matching}
C^{j}\left(s_2\right)\approx\Im_{\text{sc}}\left(\aleph^{j}\right)\cdot\Re\left(\aleph^{j}\right)\cdot C\left(s_1\right)\cdot\Re^{\text{T}}\left(\aleph^{j}\right)\cdot\Im^{\text{T}}_{\text{sc}}\left(\aleph^{j}\right)\,.
\end{equation}

The matrix $C^{j}\left(s_2\right)$ contains the periodic beam moments at the entrance to the channel and $\Im_{\text{sc}}\left(\aleph^{j}\right)$ is the periodic transport matrix of the cell including current and coupling. Since all $\Im_{\text{sc}}\left(\aleph^{n}\right)$ are products from symplectic slice matrices, all matrices $C^{n}\left(s_2\right)$ have the same eigen-emittances.

\begin{figure}[hbt]
	\centering
	\includegraphics*[width=85mm,clip=]{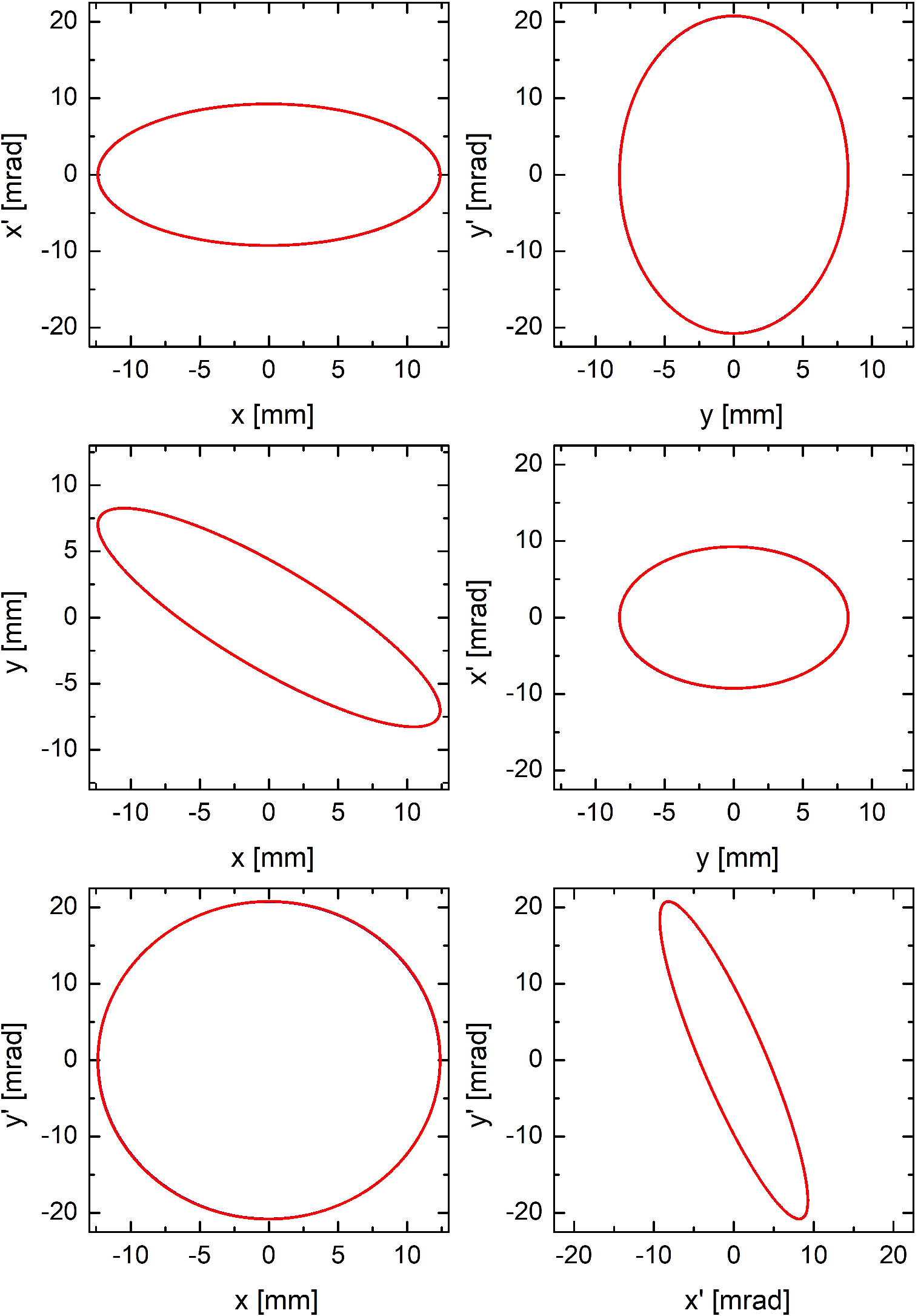}
	\caption{Projected rms-ellipses of the beam second moments matrix at the entrance (blue) and exit (red) of the first cell of the periodic channel for a coupled proton beam with~10~mA.}                                                                                                    
	\label{compare_2}
\end{figure}

In case of the example presented here, sufficient convergence has been reached at $j$~=~$4$ and the corresponding beam matrix (in units of mm and mrad) is 
\begin{equation}
\label{matched_C}
	C^{4}\left(s_2\right)=
\begin{bmatrix}
		+153.0 &-0.004 &  -86.70 &+0.006\\
		-0.004 & +85.92 & -0.004 & -170.2 \\
		  -86.70 &  -0.004 & +68.44& +0.019 \\
		+0.006 & -170.2 &+0.019 & +431.3 \\
\end{bmatrix}\,,
\end{equation}
with $\varepsilon_{x}$~=~114.6, $\varepsilon_{y}$~=~171.8, $\varepsilon_{1}$~=~268.0, and $\varepsilon_{2}$~=~18.23~mm~mrad with coupling factor of $t$~=~3.031. The corresponding output beam matrix is
\begin{equation}
	C^{4}\left(s_2+\ell\right)=
\begin{bmatrix}
		+153.3 & +0.110 &  -86.72 &+0.660\\
		+0.110 & +85.71 &-0.215 & -170.2 \\
		-86.72 & -0.215 & +68.30 & -0.131 \\
		+0.660 & -170.2 &-0.131 & +432.2 \\
\end{bmatrix}\,,
\end{equation}
the according transport matrix along the channel (one cell) is determined as
\begin{equation}
\Im_{\text{sc}}\left(\aleph^{4}\right)=
\begin{bmatrix}
		+0.476 & +1.263 & +0.126 & +0.022\\
		-0.611 & +0.476 & +0.128 & +0.038 \\
		+0.038 & +0.022 & +0.440 & +0.374 \\
		+0.128 &  +0.126 & -2.148 & +0.441 \\
\end{bmatrix}\,,
\end{equation}
with corresponding phase advances of~$\mu_{x}$~=~61.59$^{\circ}$ and $\mu_{y}$~=~63.87$^{\circ}$, respectively. Accordingly, the averaged transverse phase advance depression w.r.t. the zero-current case is~12.0\%.
Figure~\ref{compare_2} compares the six 2D-projections of the 4D-phase space ellipses $C^{4}\left(s_2\right)$ and $C^{4}\left({s_2+\ell}\right)$ in front of and behind the cell. It reveals that cell-to-cell periodicity has been achieved for all ten rms-moments of the beam matrix.

\begin{figure*}[hbt]
	\centering
	\includegraphics[width=135mm,clip=]{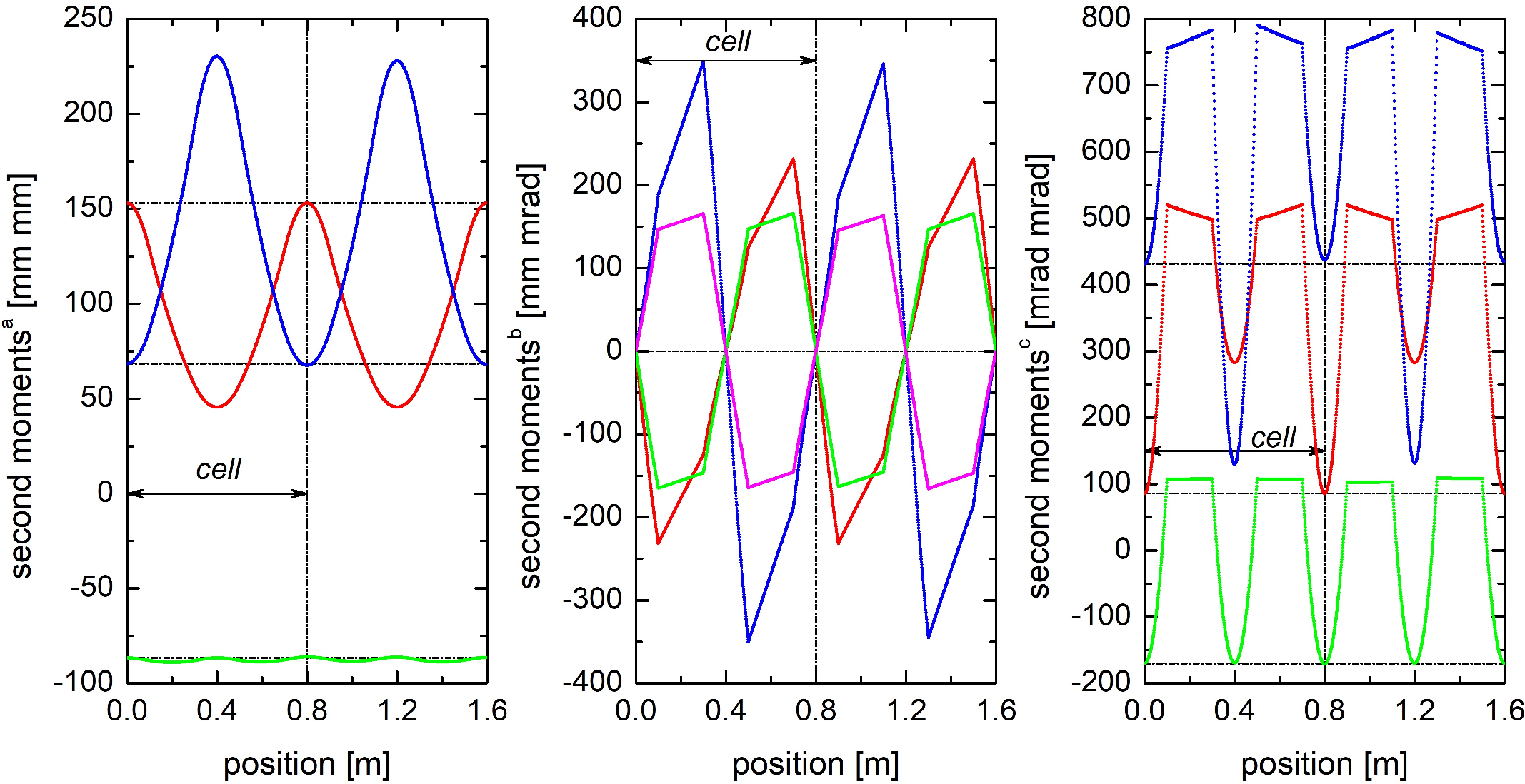}
	\caption{The ten independent rms-moments along the regular quadrupole channel (two cells) for a coupled proton beam with~10~mA. Left: rms-moments $\langle xx\rangle$, $\langle yy\rangle$, and $\langle xy\rangle$ (red, blue, and green dots); Middle: rms-moments $\langle xx'\rangle$, $\langle yy'\rangle$, $\langle xy'\rangle$, and $\langle x'y\rangle$ (red, blue, green, and magenta dots); Right: rms-moments $\langle x'x'\rangle$, $\langle y'y'\rangle$, and $\langle x'y'\rangle$ (red, blue, and green dots).}
	\label{moments_with_SC_rematching}
\end{figure*}

\begin{figure}[hbt]
	\centering
	\includegraphics*[width=85mm,clip=]{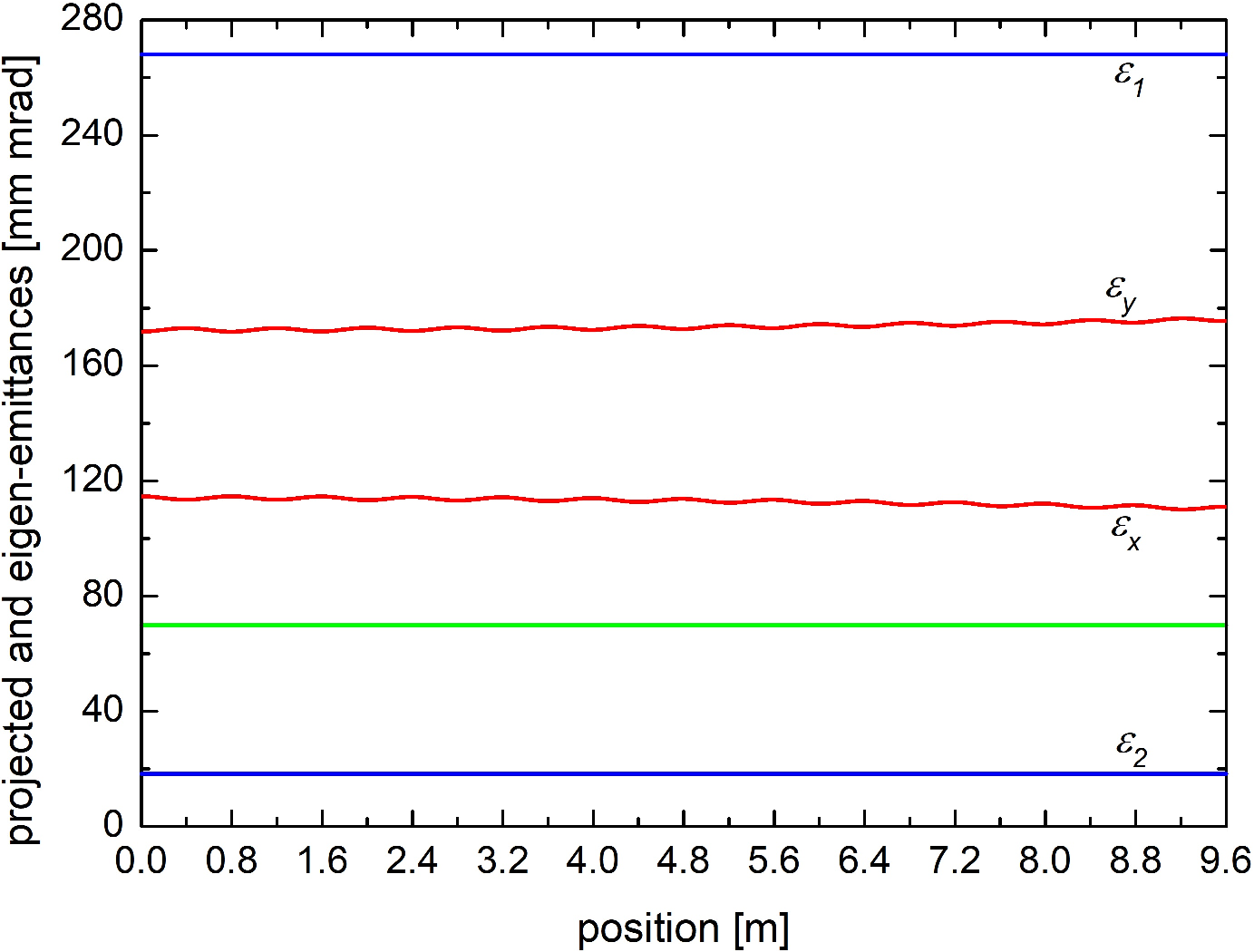}
	\caption{Projected rms-emittances (red), eigen-emittances (blue), and square root of 4d-rms-emittance (green) of a coupled 10~mA proton beam along a regular FODO quadrupole channel (twelve cells).}
	\label{emittances_B_is_on}
\end{figure}

The corresponding rms-moments along a channel comprising two cells are plotted in~Fig.~\ref{moments_with_SC_rematching}. It has been shown that cell-to-cell periodicity of an intense coupled coasting beam can be achieved under the assumption of a KV-distribution. Introduction of coupling artificially increases the projected transverse emittances. However, this growth is not intrinsic since it can be removed afterwards by decoupling. For instance, dispersive sections are parts of many beam lines, albeit they come along with horizontal emittance growth. Figure~\ref{emittances_B_is_on} plots the behaviors of 4d-rms-emittance, eigen-emittances, and projected rms-emittances along the periodic channel. However, mitigation of increase or growth of projected emittances is not the aim of the presented study. It is provision of a fully 4d-periodic solution for intense and coupled beams. In case that the solenoid field is off, all emittances remain constant ($\sqrt{\varepsilon_{1}\varepsilon_{2}}$~=~69.90~mm~mrad) along the periodic channel.

In the following chapter, the previous results shall be benchmarked with a beam featuring a Gaussian distribution. Corresponding comparisons have been done extensively for uncoupled beams during the last decades. This shall be done here for the present example to validate the method for coupled beams.
\section{Benchmarking}
Benchmarking has been done with the BEAMPATH code~\cite{BEAMPATH} using a Gaussian-type beam. The initial distribution of 2$\times$10$^4$ particles is rms-equivalent to the second beam moments matrix $C^{4}\left(s_2\right)$ from Eq.~(\ref{matched_C}). 

\begin{figure}[hbt]
	\centering
	\includegraphics*[width=85mm,clip=]{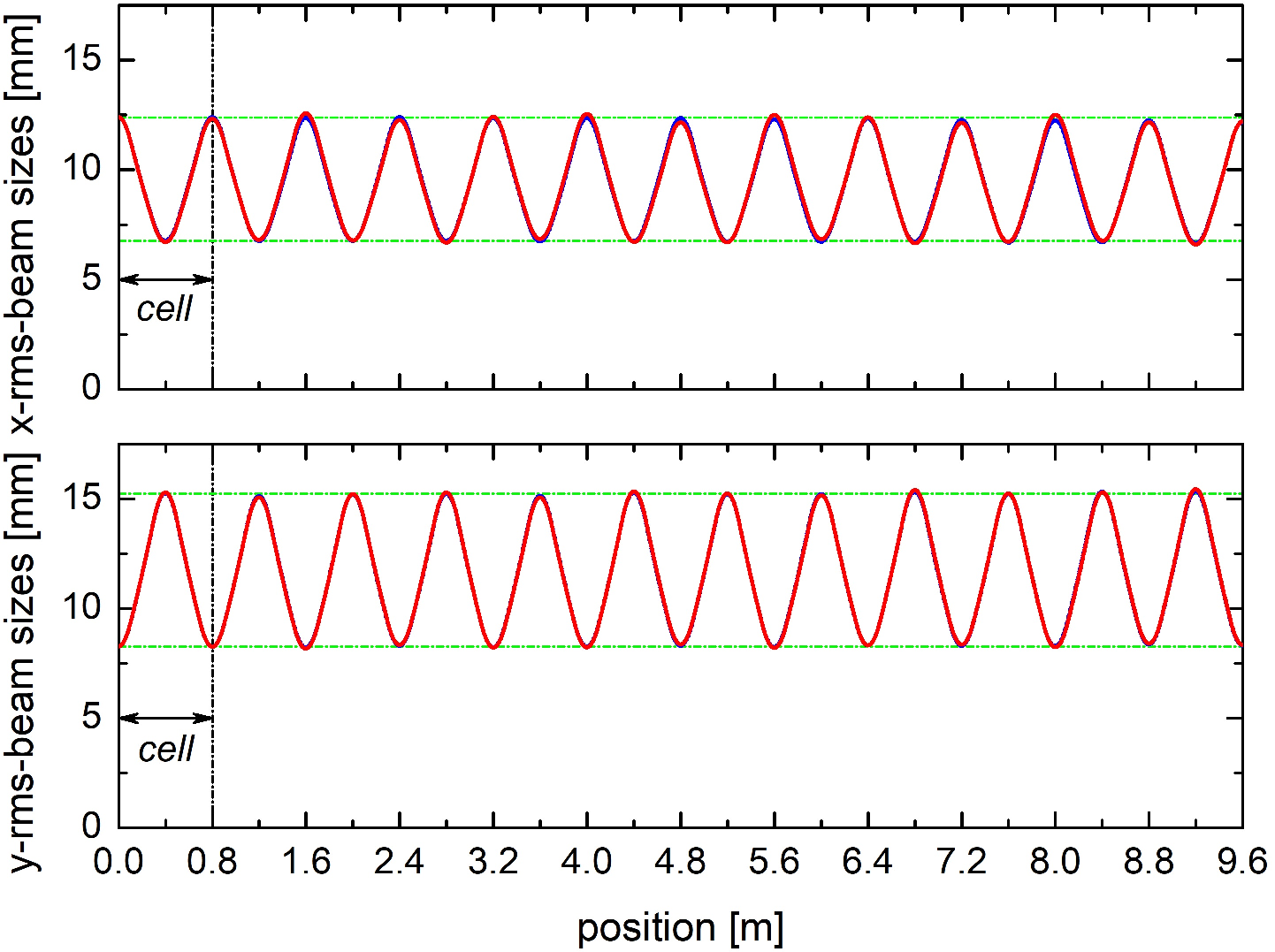}
	\caption{Transverse rms-beam sizes of a coupled 10~mA proton beam along a regular FODO quadrupole channel (twelve cells) as obtained from rms-tracking (blue) and extracted from BEAMPATH particle-tracking simulations~(red).}
	\label{benchmark_nonzero_current}
\end{figure}

Tracking has been done using 10~mA and sixty-six cells of the periodic channel. Figure~\ref{benchmark_nonzero_current} shows the transverse rms-beam sizes along the quadrupole channel obtained from the tracking method described in subsection~\ref{subsec_4b} and extracted from the simulations with BEAMPATH.

Both rms-beam sizes, from rms-tracking a KV-distribution and from simulating a Gaussian beam, reveal a high degree of matching to the lattice periodicity. The KV-based rms-beam size is perfectly regular and the Gaussian rms-beam size shows slight fluctuation around it. Some deviations are to be expected, since space charge forces especially at the outer parts of the beam are different for KV and Gaussian distributions. The matching proofed to work very well even for the Gaussian beam and a large number of cells.
\section{Conclusion}
It has been shown that cell-to-cell 4D-matching can be achieved for a coupled beam with considerable space charge forces. This has been accomplished by rms-tracking of coupled beams with KV-distribution combined with a dedicated iterative procedure of tracking and re-matching. Benchmarking with an initial Gaussian distribution along a channel with large cell number revealed that the method works very well. Hence, it provides a tool for systematic investigations of intense, coupled beam transport along periodic lattices. One special application is imposing well defined spinning to beams being transported along such lattices as drift tube linacs for instance.
\section*{Data Availability Statement}
The data that support the findings of this study are available from the corresponding author upon reasonable request.
\appendix
\section{Transfer matrices of matching section}
For zero current beam injection into the channel, the initial transfer matrix of the matching section has been assumed randomly as
\begin{equation}
	\Re\left(\aleph^{i}\right)=
\begin{bmatrix}
		+0.603 & -0.157 &  +0.001 &-0.001\\
		+0.320& +1.555 &+0.001& +0.501\\
		+0.004 & -0.024 & +0.978 & -1.686 \\
		+0.039 & +0.143 &+0.000 & +1.082 \\
\end{bmatrix}\,,
\end{equation}
satisfying Eq.~(\ref{zero_current_mismatching}). Applying the matching method (routine $Minerr$ of the MATHCAD) delivers
\begin{equation}
	\Re\left(\aleph^{0}\right)=
\begin{bmatrix}
		+0.770 & +0.402 &  -0.522 &-0.513\\
		-0.304 & +0.527 &+0.444 & -0.470 \\
		+0.352 & -0.232 & +0.427 & -0.193 \\
		+0.637 & +0.924 &+0.252 & +1.121 \\
\end{bmatrix}\,,
\end{equation}
satisfying Eq.~(\ref{zero_current_matching}). 

For high current injection, $\Re\left(\aleph^{0}\right)$ has been used as initial transfer matrix and the optimization routine has been applied again giving
\begin{equation}
	\Re\left(\aleph^{4}\right)=
\begin{bmatrix}
		+0.687 & +0.430 &  -0.400 &-0.794\\
		-0.334 & +0.371 &+0.560 & -0.393 \\
		+0.360 & -0.351& +0.379 & -0.237 \\
		+0.913 & +0.782 &+0.263 & +0.886 \\
\end{bmatrix}\,,
\end{equation}
satisfying Eq.~(\ref{high_current_matching}).
 
The field strength of the half solenoid has been used to control the coupling parameter. If the solenoid field is off, the eigen-emittances and projected emittances are equal to each other. 
Applying the identical matching routine, the transport matrices of the matching section for zero and for high current injections are obtained as 
\begin{equation}
	\Re_{\diamond}\left(\aleph^{0}\right)=
\begin{bmatrix}
		+0.431 & +1.049 &  -0.184 &+0.975\\
		-0.399 & +0.550 &-0.353 & -0.001 \\
		+0.242 & +0.230& -0.038 & -0.637 \\
		-0.255 & +1.180 &+0.972 & -0.934 \\
\end{bmatrix}\,,
\end{equation}
\begin{equation}
	\Re_{\diamond}\left(\aleph^{4}\right)=
\begin{bmatrix}
		+0.468 & +1.093 &  -0.194 &+1.028\\
		-0.370 & +0.539 &-0.333 & -0.001 \\
		+0.258 & +0.260& -0.022 & -0.706 \\
		-0.252 & +1.076 &+0.907 & -0.786 \\
\end{bmatrix}\,.
\end{equation}
\section{Uncoupled beam through the channel}
\begin{figure}[hbt]
	\centering
	\includegraphics*[width=85mm,clip=]{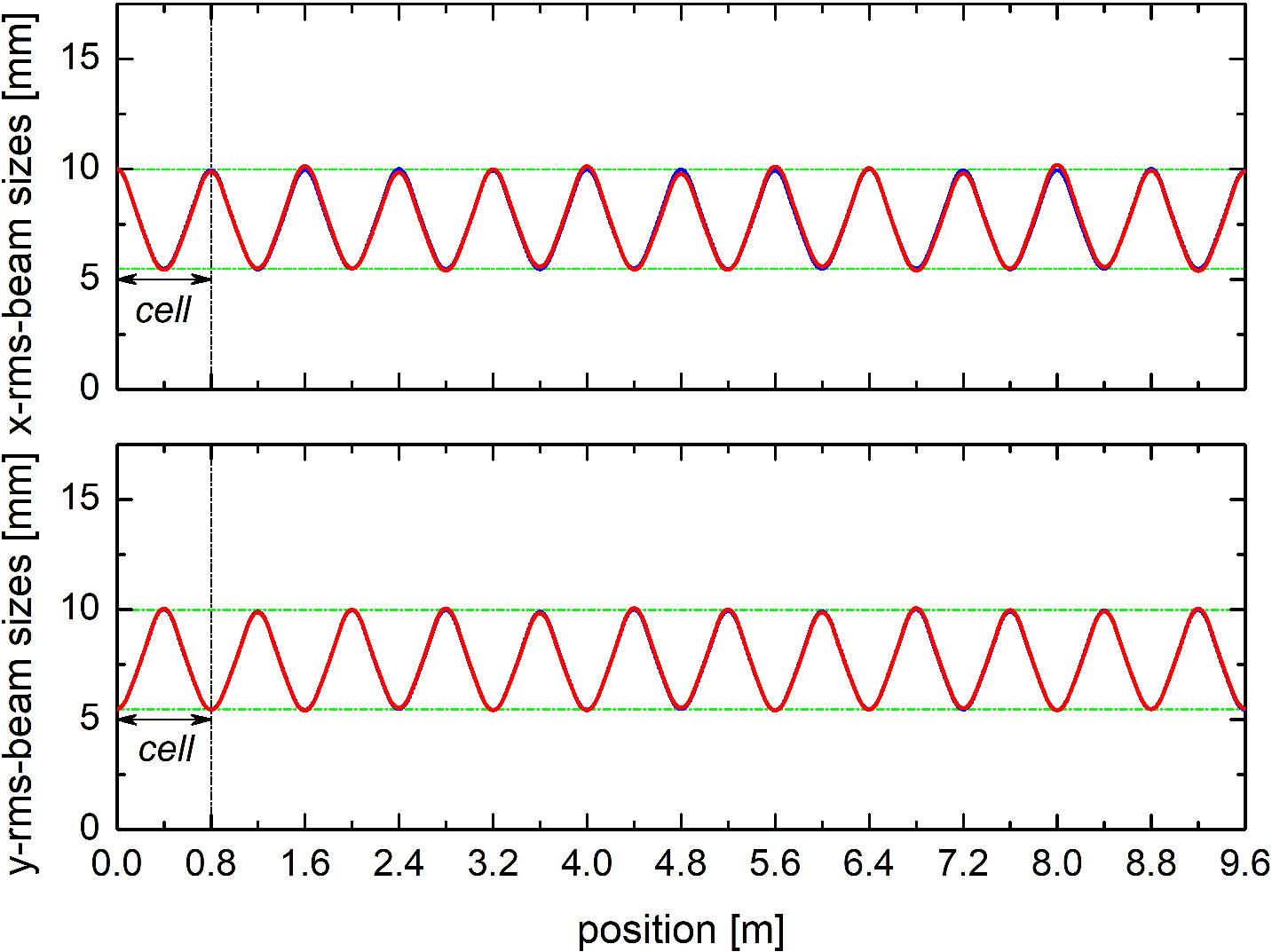}
	\caption{Transverse rms-beam sizes of a uncoupled 10~mA proton beam 
	along a regular FODO quadrupole channel (twelve cells) as obtained from rms-tracking (blue) and extracted from BEAMPATH particle-tracking simulations~(red).}
	\label{envelopes_for_uncoupled_beam}
\end{figure}

If the solenoid field is set to zero (treated as drift), the beam matrix at position $s_1$ is written as  
\begin{equation}
	C_{\diamond}\left(s_1\right)=
\begin{bmatrix}
		+133.4 & -8.192 &  +0.000 &+0.000\\
		-8.192 & +37.14 &+0.000 & +0.000 \\
		+0.000 & +0.000& +151.8 & +28.62 \\
		+0.000 & +0.000 &+28.62 & +37.60 \\
\end{bmatrix}\,.
\end{equation}
Together with the transport matrix $\Re_{\diamond}\left(\aleph^{4}\right)$, the corresponding beam matrices $C_{\diamond}^{4}\left(s_2\right)$ and $C^{4}_{\diamond}\left(s_2+\ell\right)$ are determined as
\begin{equation}
	C^4_{\diamond}\left(s_2\right)=
\begin{bmatrix}
		+99.21 & +0.011 &  +0.000 &+0.000\\
		+0.011 & +49.25 &+0.000 & +0.000 \\
		+0.000 & +0.000& +29.95 & +0.009 \\
		+0.000 & +0.000 &+0.009 & +163.2 \\
\end{bmatrix}\,
\end{equation}
and
\begin{equation}
	C^4_{\diamond}\left(s_2+\ell\right)=
\begin{bmatrix}
		+99.22 & +0.013 &  +0.000 &+0.000\\
		+0.013 & +49.25 &+0.000 & +0.000 \\
		+0.000 & +0.000& +29.95 & +0.012 \\
		+0.000 & +0.000 &+0.012 & +163.2 \\
\end{bmatrix}\,,
\end{equation}
with $\varepsilon_{x}$~=~$\varepsilon_{y}$~=~69.90~mm~mrad for both of them. The transport matrix along the channel (one cell) is determined as 
\begin{equation}
\Im^{\diamond}_{\text{sc}}\left(\aleph^{4}\right)=
\begin{bmatrix}
		+0.463 & +1.258& +0.000 & +0.000\\
		-0.642 & +0.463 & +0.000 & +0.000 \\
		+0.000 & +0.000& +0.463 & +0.380 \\
		+0.000 &  +0.000 & -2.069 & +0.463 \\
\end{bmatrix}\,,
\end{equation}
with phase advances of $\mu^{\diamond}_x$~=~62.41$^\circ$ and $\mu^{\diamond}_y$~=~62.42$^\circ$. The rms-beam sizes along the channel are plotted in Fig.~\ref{envelopes_for_uncoupled_beam}.
\section{Preliminary design of matching section}
\begin{figure}[hbt]
	\centering
	\includegraphics*[width=85mm,clip=]{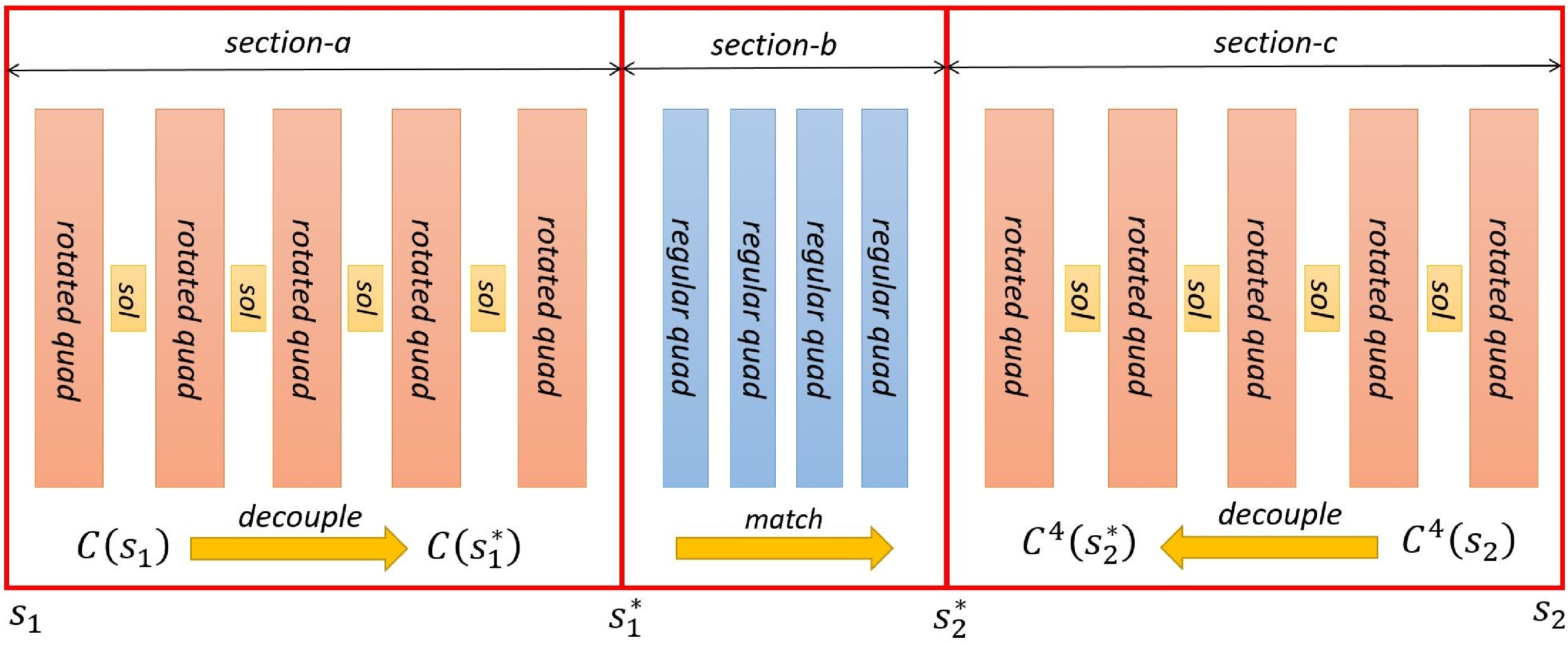}
	\caption{Conceptual matching beam line including rotated quadrupoles and solenoids. Strengths and effective lengths of the solenoid are set to $B_s$~=~0.1~T and $L$~=~0.25~m. The drift length between solenoids and rotated quadrupoles is~0.05~m.}
	\label{beamline}
\end{figure}

As mentioned previously, detailed provision of the 4D-matching beam line with space charge is a hard task being beyond the scope of this paper. However, this section shall sketch a conceptual approach to obtain an according layout. It is drawn schematically in Fig.~\ref{beamline} and it comprises three sections.
Sections~$a$ and $c$ each comprise five rotated quadrupoles being separated by solenoids. Within these sections the beam is coupled. The section~$b$ in between comprises just four regular quadrupoles, and the beam along this section is fully decoupled.

The provision of the full matching beam line starts with determination of the settings of section~$c$. It uses the known periodic solution with space charge at the beginning of the periodic channel at position~$s_2$. Its according beam moments matrix $C^4\left(s_2\right)$ is transported backwards to position~$s_2^*$. This backward transportation is done such that the resulting beam is fully decoupled at~$s_2^*$. The required settings are denoted as~$\S^c$ and they comprise the quadrupole strengths, rotation angles, and solenoid strengths. These parameters are obtained through an appropriate numerical routine ({\it Minimize} of MATHCAD for instance). The according backward transport matrix is denoted as~$\Re^{-1}_c$.

Within the second step, the settings of section~$a$ are determined numerically in order to decouple the beam at the effective half solenoid's exit at position~$s_1$. The according transport matrix is denoted as~$\Re_a$ at it provides for decoupled beam at position~$s_1^*$. Its settings are summarized as~$\S^a$.

Finally, the matching beam line is completed by an appropriate section~$b$ modeled by the transport matrix~$\Re_b$, that just provides for the matching between the two uncoupled beam matrices at~$s_1^*$ and~$s_2^*$. The transport matrix of the complete matching line hence reads as
\begin{equation}
	\Re=\Re_c\cdot\Re_b\cdot\Re_a\,.
\end{equation}
The maximum strength of the regular (rotated) quadrupoles is about~1~T/m. 
In the following the individual transport matrices are stated explicitly 
\begin{equation}
\Re_a=
\begin{bmatrix}
		+0.804 & +0.615 &  -0.797 &+0.810\\
		+0.375 & +0.874&-0.989 & +0.343 \\
		-1.078 & -1.421& -1.518 & +0.940 \\
		+1.760 & +1.832 &+2.018 & -1.562 \\
\end{bmatrix}\,,
\end{equation}
\begin{equation}
	\Re_b=
	\begin{bmatrix}
		-2.182& +1.230 &  +0.000 & +0.000\\
		+0.551 & -0.769& +0.000 & +0.000 \\
		+0.000 & +0.000 & +2.131 & +1.047 \\
		+0.000 & +0.000 &+1.658 & +1.283\\
	\end{bmatrix}\,,
\end{equation}
\begin{equation}
\Re_c=
\begin{bmatrix}
		-0.048 & +1.237 &  +0.469 &+1.417\\
		-0.296 & -0.398&-0.238 & +0.591 \\
		+0.034 & -0.857& +0.204 & +0.616 \\
		+0.681 & +0.916 &-0.344 & +0.853 \\
\end{bmatrix}\,.
\end{equation}

\begin{figure}[hbt]
	\centering
	\includegraphics*[width=85mm,clip=]{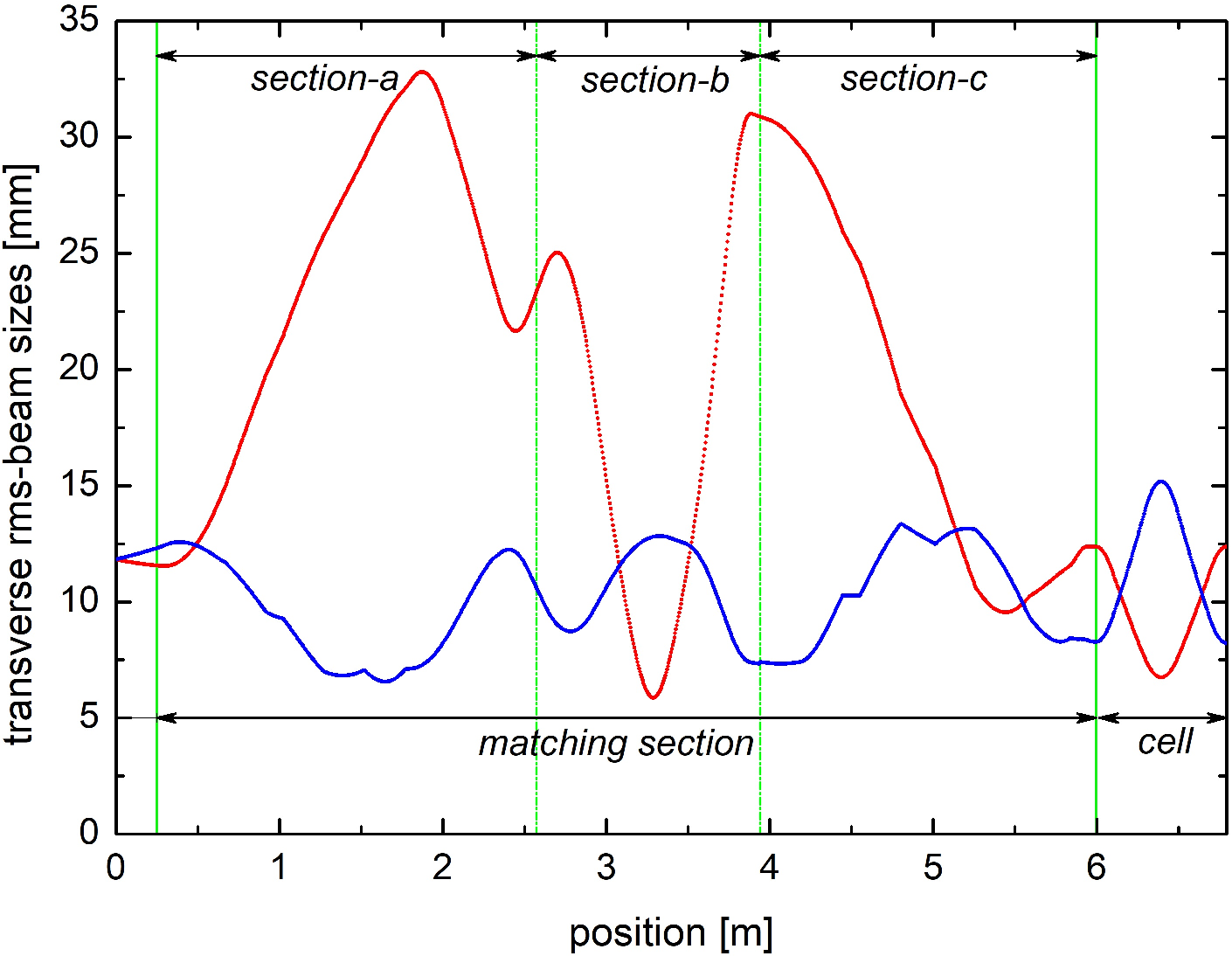}
	\caption{Transverse rms-beam sizes along half solenoid, matching section, and the first cell of a regular quadrupole FODO channel.}
	\label{end_to_end_envelopes}
\end{figure}

Corresponding transverse rms-beam sizes from positions $s_0$ to $s_2+\ell$ are shown in Fig.~\ref{end_to_end_envelopes}.

\end{document}